\documentclass{ifacconf}

\usepackage{graphicx}      
\usepackage{natbib}        
\usepackage[absolute]{textpos} 

\usepackage{amsmath}
\usepackage{amssymb}  

\providecommand{\openbox}{\leavevmode
	\hbox to.77778em{%
		\hfil\vrule
		\vbox to.675em{\hrule width.6em\vfil\hrule}%
		\vrule\hfil}}
\makeatletter
\DeclareRobustCommand{\qed}{%
	\ifmmode
	\eqno \def\@badmath{$$}
	\let\eqno\relax \let\leqno\relax \let\veqno\relax
	\hbox{\openbox}%
	\else
	\leavevmode\unskip\penalty9999 \hbox{}\nobreak\hfill
	\quad\hbox{\openbox}%
	\fi
}
\makeatother

\newtheorem{theorem}{Theorem}
\newtheorem{lemma}[theorem]{Lemma}

\newtheorem{definition}[theorem]{Definition}
\newtheorem{remark}[theorem]{Remark}
\newtheorem{assumption}[theorem]{Assumption}

\DeclareMathOperator*{\minimize}{minimize}

\begin{document}
\begin{frontmatter}

\title{Robust Stability of Suboptimal Moving Horizon Estimation using an Observer-Based Candidate Solution} 

\thanks{This work was supported by the German Research Foundation (DFG) under the research grant MU 3929/2-1.}

\author[First]{Julian D. Schiller} 
\author[Second]{Sven Knüfer} 
\author[First]{Matthias A. Müller}

\address[First]{Institute of Automatic Control, Leibniz University Hannover, Germany (e-mail:  \{schiller, mueller\}@ irt.uni-hannover.de)}
\address[Second]{Robert Bosch GmbH,
	Driver Assistance, Stuttgart, Germany (e-mail: knuefer@gmx.de)}

\begin{abstract}                
In this paper, we propose a suboptimal moving horizon estimator for nonlinear systems.
For the stability analysis we transfer the ``feasibility-implies-stability/robustness'' paradigm from model predictive control to the context of moving horizon estimation in the following sense: 
Using a suitably defined, feasible candidate solution based on the trajectory of an auxiliary observer, robust stability of the proposed suboptimal estimator is inherited independently of the horizon length and even if no optimization is performed.
\end{abstract}


\end{frontmatter}

\begin{textblock*}{\textwidth}(1.5cm,29.3cm)
	\small{
		© 2021 The Authors. This work has been accepted to IFAC for publication under a Creative Commons Licence CC-BY-NC-ND.\\
		10.1016/j.ifacol.2021.08.549
	}
\end{textblock*}

\section{Introduction}
Knowing the internal state of a dynamical system is crucial to solve many control problems, e.g. stabilizing the system via state feedback. In most practical
cases, however, the state cannot be completely measured and therefore must be reconstructed using the (measurable) system output. When considering nonlinear
systems with constraints, moving horizon estimation (MHE) has proven to be a
powerful solution to the state estimation problem, and various theoretical guarantees such as robust stability properties have been established in recent years \citep{Rawlings2017,Mueller2017,Hu2017,Knuefer2018,Allan2019a}.
In this method, the current state is estimated by optimization over a fixed number of past measurements, taking into account both system dynamics and constrained sets of decision variables. 
Since this approach requires the solution of a (usually non-convex)
optimization problem in each time step, MHE is computationally demanding and hence might not be real-time capable. 

To simplify the optimization problem, methods were developed integrating an auxiliary observer into MHE.
\cite{Sui2010}, e.g., proposed a pre-estimating MHE scheme for linear systems and used an observer to replace the state equation as a dynamical constraint.
Since this allows to compensate for model errors without computing an optimal disturbance sequence, the optimization variables could be reduced to one, namely the initial state at the beginning of the horizon.
In \cite{Suwantong2014}, this idea was transferred to a class of nonlinear systems, and a major speed improvement compared to standard MHE could be shown.
However, this results in a loss of degrees of freedom, since there is no possibility to weight model disturbances and measurement noise differently in the optimization problem.
In \cite{Liu2013}, an observer was employed to construct a confidence region for the actual system state.
Introducing this region as an additional constraint in the MHE optimization problem can, however, be quite restrictive and hence might not allow for major improvements of MHE compared to the auxiliary observer trajectory.
\cite{Gharbi2020} proposed the concept of proximity-MHE, where an observer is used to construct a stabilizing a priori estimate, yielding a proper initial guess for the optimization problem.
Considering a special class of nonlinear systems without disturbances, stability could be shown by Lyapunov arguments. 

However, all the methods discussed above still require optimal solutions to the MHE problem that need to be provided within fixed time intervals, which is in general difficult (if not impossible) to ensure.
A more intuitive solution is hence to simply terminate the underlying optimization algorithm after a fixed number of iterations, which on the one hand provides only suboptimal estimates but on the other hand ensures fixed computation-time.
However, since most results from nonlinear MHE literature are crucially based on optimality \citep[cf.][]{Rawlings2017, Mueller2017, Hu2017, Knuefer2018, Allan2019a}, stability of suboptimal MHE can not be guaranteed immediately.
For practical (real-time) applications, it is therefore crucial to develop suboptimal schemes that guarantee robust stability without requiring optimal solutions. 
Using specific optimization algorithms (gradient, conjugate-gradient, Newton-based), there are some results in the literature in this regard \citep[e.g.][]{Kuehl2011, Wynn2014, Alessandri2017, Wan2017}.
However, these rely on (local) contraction properties of the specific optimization algorithms involved and therefore require both a proper initial guess and at least one iteration to ensure (local) stability.
Recently, \cite{Gharbi2020a} incorporated a gradient-based algorithm into the framework of linear proximity-MHE, providing a suboptimal MHE scheme for linear systems without disturbances that guarantees estimator stability even if no optimization is performed. 

In this paper, we establish the ``fea\-si\-bil\-i\-ty-im\-plies-sta\-bil\-i\-ty/ro\-bust\-ness'' paradigm from model predictive control (MPC) in the context of nonlinear MHE.
In MPC it holds that if the optimization provides a feasible, suboptimal solution that improves the cost of a well-chosen warm start, (robust) stability of the controller is inherited \cite[cf.][]{Scokaert1999,Pannocchia2011a}.
Applying this concept to MHE, we establish robust stability of the proposed suboptimal moving horizon estimator by requiring that a suboptimal solution is (i) feasible and (ii) improves the cost of a candidate solution, which we subsequently construct using an auxiliary observer.
In particular, we consider exponentially detectable nonlinear systems \cite[cf.][]{Knuefer2018,Knuefer2020,Allan2020}, and together with an exponentially convergent auxiliary observer we show robust exponential stability of the proposed moving horizon estimator independently of the horizon length and even if no optimization is performed.
Note that in \cite{Liu2013}, the observer trajectory could also be used as a candidate solution potentially yielding a similar result, but here we consider the most general (and practically relevant) MHE problem formulation without additional (potentially restrictive) constraints, and we also allow for a general cost function in the MHE optimization problem without the need for certain additional terms and/or specific weighting constants as was, e.g., required in \cite{Hu2017}, \cite{Mueller2017} and \cite{Allan2019a} to establish robust stability of MHE.

The paper is organized as follows. In Section~\ref{sec:problem}, the proposed suboptimal estimator together with the candidate solution and assumptions regarding system and observer are stated. 
Robust stability of the estimator is shown in Section~\ref{sec:RGES} and, to conclude this paper, an illustrative simulation example is presented in Section~\ref{sec:simulation}. 

\textit{Notation:}
Let the set of all integers in an interval $[a,b] \subset \mathbb{R}$ be denoted by $\mathbb{I}_{[a,b]}$ and the
set of all integers greater than or equal to $a$ by $\mathbb{I}_{\geq a}$.
We define $|x|$ to be the Euclidean norm of the vector $x \in \mathbb{R}^n$. 
Symbols in bold type represent sequences of vectors, i.e. $\boldsymbol{x} = \lbrace x(0), x(1),...\rbrace$.
A function $\alpha:\mathbb{R}_{\geq0}\rightarrow \mathbb{R}_{\geq0}$ is of class $\mathcal{K}$ if it is continuous, strictly increasing and $\alpha(0) = 0$. 
If in addition $\lim_{s\rightarrow \infty}\alpha(s) = \infty$, it is of class $\mathcal{K}_{\infty}$.
For every $\alpha \in \mathcal{K}$, we have that
\begin{equation}
\alpha(a_1 + ... + a_n) \leq \alpha(na_1) + ... + \alpha(na_n) \label{eq:Kprop}
\end{equation} 	
for all $a_i \in \mathbb{R}_{\geq0}$ with $i \in \mathbb{I}_{[1,n]}$ \citep[for a proof, see, e.g.,][Appendix A]{Rawlings2012}.

\section{PROBLEM SETUP} \label{sec:problem}


\subsection{System description and basic definitions}
We consider nonlinear discrete-time systems of the form
\begin{equation}\label{eq:system} 
\begin{split}
x(t+1) &= f(x(t)) + w(t), \qquad x(0) = x_0, \\ y(t) &= h(x(t)) + v(t),
\end{split}
\end{equation}
where $t \in \mathbb{I}_{\geq0}$ and where $x \in \mathbb{X} \subseteq \mathbb{R}^n$ is the system state, $y \in \mathbb{Y} \subseteq  \mathbb{R}^p$ is the measured output, $w \in \mathbb{W} \subseteq \mathbb{R}^n$ is the (unknown) process disturbance, and $v \in \mathbb{V} \subseteq  \mathbb{R}^p$ is the (unknown) measurement noise.
We assume that the sets $\mathbb{X}$, $\mathbb{Y}$, $\mathbb{W}$, and $\mathbb{V}$ are closed and nonempty.
Furthermore, $f:\mathbb{X} \rightarrow \mathbb{X}$ and $h: \mathbb{X} \rightarrow \mathbb{Y}$ are nonlinear continuous functions, and we additionally impose the following assumption on~$h$.

\begin{assumption} \label{ass:lipschitz}
	The function $h: \mathbb{X} \rightarrow \mathbb{Y}$ is Lipschitz continuous in $\mathbb{X}$ with Lipschitz constant $L_h$.
\end{assumption} 

The solution to system (\ref{eq:system}) at time $t \in \mathbb{I}_{\geq0}$, starting at the initial condition $x_0$ and being driven by the disturbance sequence $\boldsymbol{w}(t)=\lbrace w(0),\dots,w(t-1) \rbrace$, is denoted by $x(t;x_0,\boldsymbol{w}(t))$, or just by $x(t)$ if there is no ambiguity about $x_0$ and $\boldsymbol{w}(t)$.
In the following, we consider the state estimation problem for system (\ref{eq:system}).
Motivated by \cite{Allan2020}, we define a general state estimator as a sequence of maps.


\begin{definition}[State Estimator]
	A state estimator for system (\ref{eq:system}) is a sequence of functions $\Psi_t:\mathbb{X}\times \mathbb{Y}^{t}\rightarrow \mathbb{X}$ that produces for all $t \in \mathbb{I}_{\geq0}$ a state estimate
	\begin{equation} \label{eq:state_estimator}
	z(t)=\Psi_t(z_0,\boldsymbol{y}(t)),
	\end{equation}
	where $z_0$ is an initial estimate and $\boldsymbol{y}(t)= \lbrace y(0),\dots,y(t-1) \rbrace$ is a sequence of measured outputs.
\end{definition}

Note that this definition applies to full-order state observers as well as to more advanced approaches, e.g., full information and moving horizon estimators \cite[cf.][]{Allan2020}. 
To appropriately describe robust stability of a general state estimator in the form of (\ref{eq:state_estimator}), we employ the following notion.

\begin{definition}[RGES] \label{def:RGES}
	A state estimator for system (\ref{eq:system}) in the form of (\ref{eq:state_estimator}) is called \textit{robustly globally exponentially stable} (RGES) if there exist constants $C_p,C_w,C_v>0$ and $\rho \in (0,1)$ such that for all initial conditions $x_0,z_0 \in \mathbb{X}$ and all disturbance sequences $\boldsymbol{w}(t)$ and $\boldsymbol{v}(t)$ with $w(i) \in \mathbb{W}$, $v(i) \in \mathbb{V}$ for all $i \in \mathbb{I}_{[0,t-1]}$, the following holds for all $t \in \mathbb{I}_{\geq0}$: 
	\begin{align}\label{eq:RGES} 
	| x(t&) - z(t) | \leq
	C_p|x_0 - z_0|\rho^{t} \\
	& + C_w \sum_{\tau=1}^{t}\rho^{\tau}|w(t-\tau)|
	+ C_v \sum_{\tau=1}^{t}\rho^{\tau}|v(t-\tau)|. \notag
	\end{align} 
\end{definition}

This definition of robust exponential stability, including the explicit time-discounting in the disturbance terms, has recently been introduced and analyzed in the context of MHE \citep{Knuefer2018,Knuefer2020,Allan2020}. One advantage of this formulation is that it directly reveals that the estimation error (exponentially) converges to zero if both $w$ and $v$ converge to zero. 
In order to establish RGES of the proposed estimator, we will exploit the following exponential detectability condition.

\begin{definition}[e-IOSS] \label{def:eIOSS} System (\ref{eq:system}) is  \textit{incrementally exponentially input/output-to-state stable} (e-IOSS) if there exist constants $ c_p,c_w,c_v>0$ and $\eta \in (0,1)$ such that for each pair of initial conditions $x_1,x_2 \in \mathbb{X}$ and each two disturbance sequences $\boldsymbol{w}_1(t)$ and $\boldsymbol{w}_2(t)$ with $w_1(i),w_2(i) \in \mathbb{W}$ for all $i \in \mathbb{I}_{[0, t-1]}$, the following holds for all $t \in \mathbb{I}_{\geq0}$:
	\begin{align} 
	| x_1(&t) - x_2(t) | 
	\leq  c_p|x_1 - x_2|\eta^t \notag \\
	& \ \ \ + c_w\sum_{\tau=1}^t\eta^{\tau}|w_1(t-\tau) - w_2(t-\tau)| \label{eq:eIOSS} \\
	&\ \ \ + c_v\sum_{\tau=1}^t\eta^{\tau}|h(x_1(t-\tau) - h(x_2(t-\tau))|, \notag
	\end{align}
	where $x_i(t) = x(t;x_i,\boldsymbol{w}_i(t)), i = 1,2$.
\end{definition}

Note that e-IOSS in the sense of Definition \ref{def:eIOSS} is an extension of the classical \textit{incremental input/output-to-state stability} (i-IOSS), which became standard as a notion of nonlinear detectability in the context of MHE in recent years \citep{Rawlings2017,Mueller2017,Hu2017,Allan2019a}.
However, since i-IOSS only considers the maximum norm of \textit{all} past input and output differences, asymptotic convergence of two trajectories can only be shown indirectly. 
To overcome this issue, i-IOSS was extended in \cite{Knuefer2018, Knuefer2020} and \cite{Allan2020} by explicitly discounting input and output differences over time, and its equivalence to standard i-IOSS was shown \cite[see][]{Allan2020}.
Since \textit{exponential} discounting is a special case of the more general notions considered in \cite{Knuefer2020} and \cite{Allan2020}, the same results apply: (i)~e-IOSS implies i-IOSS, (ii)~e-IOSS is necessary for the existence of an RGES estimator and (iii)~e-IOSS can be shown by using a suitable Lyapunov function.

\subsection{Suboptimal moving horizon estimator}

Given some finite estimation horizon $N \in \mathbb{I}_{\geq 1}$, the objective function of an MHE problem is traditionally defined for time $t \in \mathbb{I}_{\geq N}$ as
\begin{align} \label{eq:costfunction}
J(\chi(t-N|t)&, \boldsymbol{\omega}(t)) := \Gamma(\chi(t-N|t),\bar{x}_{t-N}) \notag \\
&+ \sum_{i=t-N}^{t-1}l(\omega(i|t),\nu(i|t)),
\end{align}
where $\chi(t-N|t)$ denotes the estimated state for time $t-N$, estimated at time $t$, and $\omega(i|t)$ and $\nu(i|t)$ denote estimates of the process disturbance and measurement noise for time~$i$, estimated at time $t$. 
Define the sequences $\boldsymbol{\omega}(t):= \lbrace \omega(t-N|t), \dots, \omega(t-1|t) \rbrace$ and $\boldsymbol{\nu}(t):= \lbrace \nu(t-N|t), \dots, \nu(t-1|t) \rbrace$ and note that $\boldsymbol{\nu}(t)$ can be expressed in terms of the decision variables $\chi(t-N|t)$ and $\boldsymbol{\omega}(t)$ by using the system dynamics (\ref{eq:system}).
Suitable choices for stage cost~$l$, prior weighting $\Gamma$ and prior $\bar{x}_{t-N}$ are specified below.
Given the $N$ past measurements $y(t-N),...,y(t-1)$, the moving horizon estimate is typically designed to be the minimizer of the following optimization problem:
\begin{equation} \label{eq:optiMHE}
\mathcal{P}_t  :  \ \minimize_{\chi(t-N|t), \boldsymbol{\omega}(t)} J(\chi(t-N|t), \boldsymbol{\omega}(t))
\end{equation}
subject to
\begin{align} 
	&\chi(i+1|t) = f(\chi(i|t)) + \omega(i|t), \notag \\
	&y(i) = h(\chi(i|t)) + \nu(i|t), \label{eq:constraints} \\
	&\chi(i|t) \in \mathbb{X}, \ \omega(i|t) \in \mathbb{W}, \ \nu(i|t) \in \mathbb{V}, \quad i \in \mathbb{I}_{[t-N,t-1]}. \notag
\end{align}

In order for $\mathcal{P}_t$ to be well-defined, we make the following assumptions about the stage cost $l$ and prior weighting $\Gamma$.

\begin{assumption} \label{ass:stagecosts}
	The prior weighting $\Gamma : \mathbb{X} \times \mathbb{X} \rightarrow \mathbb{R}_{\geq0}$ and stage cost $l : \mathbb{W} \times \mathbb{V} \rightarrow \mathbb{R}_{\geq0}$ are continuous and satisfy
	\begin{align} \label{eq:cond0}
	\begin{array}{c}
	\underline{\gamma}_p(|\chi-\bar{x}|) \leq \Gamma(\chi,\bar{x}) \leq \overline{\gamma}_p(|\chi-\bar{x}|) \\[1ex]
	\underline{\gamma}_w(|\omega|) + \underline{\gamma}_v(|\nu|) \leq l(\omega,\nu) \leq  \overline{\gamma}_w(|\omega|) + \overline{\gamma}_v(|\nu|) 
	\end{array}
	\end{align}
	for all $\chi, \bar{x} \in \mathbb{X}$, all $\omega \in \mathbb{W}$ and all $\nu \in \mathbb{V}$, where $\underline{\gamma}_p,\allowbreak \overline{\gamma}_p,\allowbreak \underline{\gamma}_w,\allowbreak \overline{\gamma}_w,\allowbreak \underline{\gamma}_v,\allowbreak \overline{\gamma}_v\allowbreak$ satisfy
	\begin{equation}\label{eq:exp_bounds}
		\underline{\gamma}_i(s) = \underline{c}_is^a, 
		\quad
		\overline{\gamma}_i(s) = \overline{c}_is^a,
		\quad i \in \lbrace p, w, v \rbrace
	\end{equation}	
	for some constants $\underline{c}_p, \overline{c}_p, \underline{c}_w, \overline{c}_w, \underline{c}_v, \overline{c}_v>0$ and $a>0$. 
\end{assumption}

Now, rather than solving $\mathcal{P}_t$ to optimality at each time $t \in \mathbb{I}_{\geq N}$, we consider the following suboptimal estimator.

\begin{definition}[Suboptimal Estimator] \label{def:estimator}
	For any $t \in \mathbb{I}_{\geq N}$, let the tupel $(\tilde{x}(t-N|t), \tilde{\boldsymbol{w}}(t))$ denote a feasible \textit{candidate solution} to the MHE problem (\ref{eq:optiMHE}).
	Then, the corresponding suboptimal solution of (\ref{eq:optiMHE}) is defined as \textit{any} tupel $(\hat{x}(t-N|t), \hat{\boldsymbol{w}}(t))$ that satisfies (i) the MHE constraints (\ref{eq:constraints}) and (ii) the cost decrease condition
	\begin{equation} \label{eq:estimator}
	J(\hat{x}(t-N|t), \hat{\boldsymbol{w}}(t)) \leq 
	J(\tilde{x}(t-N|t), \tilde{\boldsymbol{w}}(t)).
	\end{equation}
	The state estimate at time $t \in \mathbb{I}_{\geq N}$ is then defined as
	$\hat{x}(t) = x(t;\hat{x}(t-N|t),\hat{\boldsymbol{w}}(t))$.
\end{definition}

Note that (\ref{eq:estimator}) ensures that at a given time $t$, the cost of a suboptimal solution is no larger than the cost of the candidate solution.
This can be guaranteed in general by nearly all numerical solvers applied to $\mathcal{P}_t$ subject to (\ref{eq:constraints}) and (\ref{eq:estimator}), if they are initialized with the candidate solution as a warm start and then terminated after a finite number of iterations (including 0), cf. \cite{Pannocchia2011a}.
In the following, we use $\hat{J}(t)$ and $\tilde{J}(t)$ to denote the cost obtained for the suboptimal and the candidate solution at time $t \in \mathbb{I}_{\geq N}$, i.e. $\hat{J}(t) := J(\hat{x}(t-N|t), \hat{\boldsymbol{w}}(t))$ and $\tilde{J}(t) := J(\tilde{x}(t-N|t), \tilde{\boldsymbol{w}}(t))$, respectively.

\subsection{Pre-stabilizing candidate solution}

We design the required candidate solution used in Definition \ref{eq:estimator} based on some auxiliary state estimator (\ref{eq:state_estimator}).
In particular, we assume that this estimator is in fact some RGES full-order state observer in the following form.

\begin{assumption} \label{ass:observer_0}
	For system (\ref{eq:system}), there exists an RGES full-order state observer according to
	\begin{equation} \label{eq:observer}
	z(t+1) = f(z(t)) + L(z(t), y(t) - h(z(t)))
	\end{equation}
	with the state $z \in \mathbb{X}$, functions $f$ and $h$ from (\ref{eq:system}), correction term $L: \mathbb{X} \times \mathbb{Y} \rightarrow \mathbb{X}$ with $L(\cdot,0) = 0$, and such that (\ref{eq:RGES}) is satisfied for all $t \in \mathbb{I}_{\geq0}$.
\end{assumption}

If this latter assumption holds, (suboptimal) MHE can then be used in order to improve the state estimate compared to that obtained directly from the auxiliary observer; this will also be illustrated in our simulation example in Section~\ref{sec:simulation}.
Furthermore, note that in (\ref{eq:observer}) we require a full-order state observer in output injection form.  This is not restrictive, since from \cite{Knuefer2020} and \cite{Sontag1997} it follows that \textit{any} robustly stable full-order state observer must in fact have this form.
As a result, we can interpret every fitting error $v_z = y-h(z)$ and every correction term $L(z,v_z)$ directly as estimates of disturbances $v$ and $w$, respectively. 
To this end we need some further assumptions regarding the observer.

\begin{assumption} \label{ass:observer}
	The correction term $L(z,v_z)$ can be linearly bounded by its second argument, i.e. there exists some $\kappa>0$ such that
	\begin{equation} \label{eq:kappa}
	|L(\cdot, v_z(t))| \leq \kappa|v_z(t)|. 
	\end{equation}
	Moreover, $v_z(t) \in \mathbb{V}$ and $L(z(t), v_z(t)) \in \mathbb{W}$ for all $t \in \mathbb{I}_{\geq0}$.
\end{assumption}

Equation (\ref{eq:kappa}) is motivated by the observation that, for many common observers in the form of (\ref{eq:observer}), the correction term can usually be reduced to a simple (constant or time-varying) weighting of the fitting error (this is, e.g., the case for Luenberger-like observers, Kalman filter-based observers and high-gain observers).
However, assuming that $z,v_z$ and $L$ evolve in $\mathbb{X}$, $\mathbb{V}$ and $\mathbb{W}$ can be rather restrictive, but this condition is necessary to make sure that every candidate solution constructed by observer (\ref{eq:observer}) results in a feasible state trajectory of system (\ref{eq:system}).
If this latter assumption is not satisfied, one could just omit these set constraints.
Note that this does not change the theoretical guarantees derived in Section \ref{sec:RGES}, but might deteriorate the MHE performance in practice.
We define the candidate solution as follows.

\begin{definition}[Candidate Solution] \label{def:warmstart}
	For time $t \in \mathbb{I}_{\geq N}$, the candidate solution $(\tilde{x}(t-N|t), \tilde{\boldsymbol{w}}(t))$ follows the trajectory of an RGES observer in the form of (\ref{eq:observer}), satisfying Assumption~\ref{ass:observer}, i.e.
	\begin{align} \label{eq:warmstart}
	\begin{split}
	&\tilde{x}(t-N|t) = z(t-N), \\
	&\tilde{\boldsymbol{w}}(t) = \lbrace L(t-N),...,L(t-1) \rbrace,
	\end{split} 
	\end{align}
	where $L(t) = L(z(t),v_z(t))$. 
\end{definition}

Moreover, the estimated state trajectory resulting from observer~(\ref{eq:observer}) not only serves to define the candidate solution for the suboptimal MHE problem but also defines the prior according to 
\begin{equation} \label{eq:prior}
\bar{x}_{t-N} = z(t-N)
\end{equation}
for all $t \in \mathbb{I}_{\geq N}$. 
Together with the assumptions of an e-IOSS system and an RGES auxiliary observer, these last two choices made in (\ref{eq:warmstart}) and (\ref{eq:prior}) are the key steps to establish RGES of the proposed suboptimal moving horizon estimator in the following section. 

\begin{remark} \label{rem:emtpyN}
	For $t \in \mathbb{I}_{[0,N-1]}$, i.e. until the estimation horizon is full, in (\ref{eq:costfunction})-(\ref{eq:prior}) every $N$ be must replaced by~$t$.
\end{remark}


\section{Robust stability of suboptimal MHE} \label{sec:RGES}
In order to prove robust stability of the proposed suboptimal moving horizon estimator, we first need the following auxiliary lemma, which provides an upper bound for the suboptimal cost $\hat{J}(t)$.

\begin{lemma} \label{lem:boundedness_of_J}
	Suppose that system (\ref{eq:system}) is e-IOSS and that Assumptions \ref{ass:lipschitz}, \ref{ass:stagecosts}, \ref{ass:observer_0}, and \ref{ass:observer} are satisfied.	
	Choose $N\in \mathbb{I}_{\geq 1}$ and $\bar{x}_0\in\mathbb{X}$ arbitrarily and let $z_0 = \bar{x}_0$. 
	Then, for all $t \in \mathbb{I}_{\geq 0}$, the suboptimal cost $\hat{J}(t)$ evaluated at an estimate provided by the moving horizon estimator from Definition~\ref{def:estimator} using the candidate solution from (\ref{eq:warmstart}) satisfies the following: 
	\begin{align}  
	&\hat{J}(t) \leq  
	C_p^a\bar{c} \bar{\rho}_1(N)|x_0 - \bar{x}_0|^a\rho^{at} \notag \\
	& \ + C_w^a
	\bar{c}\bar{\rho}_2(N) 
	\Big(\sum_{\tau=1}^t \rho^{\tau-1}|w(t-\tau)|\Big)^a \label{eq:lem_1} \\
	& \ + C_v^a
	\bar{c} \bar{\rho}_2(N)  
	\Big(\sum_{\tau=1}^t \rho^{\tau-1}|v(t-\tau)|\Big)^a , \notag
	\end{align}
	where $\bar{c}:= (3\bar{L})^a(\overline{c}_w\kappa^a+\overline{c}_v)$, $\bar{\rho}_1(N) = (\rho^{-aN}-1)/(1-\rho^a)$ and $\bar{\rho}_2(N) = (\rho^{-a(N-1)}- \rho^{a})/(1-\rho^a)$.
\end{lemma}

\begin{pf}
	Consider observer (\ref{eq:observer}). 
	Applying (\ref{eq:system}) and Assumption \ref{ass:lipschitz}, the fitting error $v_z(t)$ can be bounded by
	\begin{equation}
	|v_z(t)| \leq L_h|x(t) - z(t)| + |v(t)|. 
	\end{equation}
	Since the observer is RGES according to Definition \ref{def:RGES}, we can insert (\ref{eq:RGES}) using $z_0 = \bar{x}_0$. 
	With $\bar{L}:= \max \lbrace L_h , 1/C_v \rbrace$ we can move $|v(t)|$ into the corresponding sum and hence obtain
	\begin{align} 
	|v_z(t)| &\leq \bar{L} \Big( C_p|x_0 - \bar{x}_0|\rho^t 
	+ C_w\sum_{\tau=0}^t \rho^{\tau}|w(t-\tau)|  \notag \\
	& \ \ \   + C_v\sum_{\tau=0}^t \rho^{\tau}|v(t-\tau)| \Big). \label{eq:proof_2}
	\end{align} 
	Note that the sums now start at $\tau = 0$.
	By Assumption~$\ref{ass:observer}$, multiplying (\ref{eq:proof_2}) by $\kappa$ yields a similar bound for $|L(t)|$.
	Now, for $t \in \mathbb{I}_{\geq N}$, consider the suboptimal moving horizon estimator from (\ref{eq:estimator}) together with the candidate solution defined in (\ref{eq:warmstart}). 
	Since $\tilde{x}(t-N|t) = \bar x_{t-N} = z(t-N)$, we have that $\Gamma(\tilde{x}(t-N|t),\bar{x}_{t-N})=0$ and thus 
	\begin{gather} 
	\hat{J}(t) \stackrel{(\ref{eq:estimator})}{\leq} 
	\tilde{J}(t)
	\stackrel{(\ref{eq:cond0}),(\ref{eq:warmstart})}{\leq} 
	\sum_{i = t-N}^{t-1}\Big( \overline{\gamma}_w(|L(i)|) + \overline{\gamma}_v(|v_z(i)|) \Big). \notag
	\end{gather}
	Applying (\ref{eq:proof_2}) and (\ref{eq:kappa}) for all $i \in \mathbb{I}_{[t-N,t-1]}$ yields 
	\begin{align} 
	&\hat{J}(t) 
	\stackrel{(\ref{eq:Kprop}),(\ref{eq:exp_bounds})}{\leq}  
	(C_p)^a\bar{c} \sum_{i = t-N}^{t-1} 
	(|x_0 - \bar{x}_0|\rho^i)^a \notag \\
	& \ \ \ \ \ \ + (C_w)^a
	\bar{c}\sum_{i = t-N}^{t-1} 
	\Big(\sum_{\tau=0}^i \rho^{\tau}|w(i-\tau)|\Big)^a \label{eq:proof_rho} \\
	& \ \ \ \ \ \ + (C_v)^a
	\bar{c} \sum_{i = t-N}^{t-1} 
	\Big(\sum_{\tau=0}^i \rho^{\tau}|v(i-\tau)|\Big)^a, \notag
	\end{align}
	where $\bar{c}:= (3\bar{L})^a(\overline{c}_w\kappa^a+\overline{c}_v)$.
	Next, we employ properties of the geometric series. First, note that
	\begin{equation} \label{eq:proof_rho4}
	\sum_{i=t-N}^{t-1}(\rho^{a})^{i} 
	= \frac{\rho^{a(t-N)}-\rho^{at}}{1-\rho^a} 
	= \bar{\rho}_1(N)\rho^{at}
	\end{equation}
	with $\bar{\rho}_1(N):= (\rho^{-aN}-1)/(1-\rho^a)$.
	Now  we expand the inner sum of the first double sum in (\ref{eq:proof_rho}) by adding additional disturbance terms from $i+1$ to $t-1$,
	\begin{align}
	\sum_{i = t-N}^{t-1} 
	\Big( &\sum_{\tau=0}^i \rho^{\tau}|w(i-\tau)| \Big) ^a \\
	&\leq 	\sum_{i = t-N}^{t-1} (\rho^a)^{i-t+1}
	\Big( \sum_{\tau=0}^{t-1} \rho^{\tau}|w(t-1-\tau)| \Big) ^a. \notag
	\end{align}
	Defining $j:=i-t+N+1$ yields
	\begin{align} \label{eq:proof_rho5}
	\sum_{i = t-N}^{t-1} (\rho^a&)^{i-t+1} 
	= \sum_{j = 1}^{N} (\rho^a)^{j-N} \\
	&= (\rho^{-a(N-1)}- \rho^{a})/(1-\rho^a) =: \bar{\rho}_2(N). \notag
	\end{align}
	An analogous result holds for $\sum_{i = t-N}^{t-1} 
	(\sum_{\tau=0}^i \rho^{\tau}|v(i-\tau)|)^a$. Applying (\ref{eq:proof_rho4})-(\ref{eq:proof_rho5}) to (\ref{eq:proof_rho}), we obtain (\ref{eq:lem_1}) for $t \in \mathbb{I}_{\geq N}$.
	Now it remains to show that (\ref{eq:lem_1}) is also satisfied for $t \in \mathbb{I}_{[0,N-1]}$.
	Recall Remark \ref{rem:emtpyN} and by applying similar arguments as before, we obtain (\ref{eq:lem_1}) with $\bar{\rho}_1(N)$ and $\bar{\rho}_2(N)$ replaced by $\bar{\rho}'_1(t) = (\rho^{-at}-1)/(1-\rho^a)$ and $\bar{\rho}'_2(t) = (\rho^{-a(t-1)}-\rho^a)/(1-\rho^a)$, respectively.
	Since $t < N$, we have both $\bar{\rho}_1(N) > \bar{\rho}'_1(N)$ and $\bar{\rho}_2(N) > \bar{\rho}'_2(N)$, and hence (\ref{eq:lem_1}) holds for all $t \in \mathbb{I}_{\geq 0}$, which completes the proof. \qed
\end{pf}

\begin{theorem} \label{theorem}
	Suppose that system (\ref{eq:system}) is e-IOSS and that Assumptions \ref{ass:lipschitz}, \ref{ass:stagecosts}, \ref{ass:observer_0} and \ref{ass:observer} are satisfied.	
	Choose $N\in \mathbb{I}_{\geq 1}$ and $\bar{x}_0\in\mathbb{X}$ arbitrarily and let $z_0 = \bar{x}_0$. 
	Then, the suboptimal moving horizon estimator from Definition \ref{def:estimator} using the candidate solution from (\ref{eq:warmstart}) is RGES, 
	i.e. there exist constants $C_1,C_2,C_3>0$ and $\lambda \in (0,1)$ such that 
	the following holds for all $t \in \mathbb{I}_{\geq0}$: 
	\begin{align}\label{eq:RGES_result} 
	| x(t&) - \hat{x}(t) | \leq
	C_1|x_0 - \bar{x}_0|\lambda^{t} \\
	& + C_2 \sum_{\tau=1}^{t}\lambda^{\tau}|w(t-\tau)|
	+ C_3 \sum_{\tau=1}^{t}\lambda^{\tau}|v(t-\tau)|. \notag
	\end{align} 
\end{theorem} 	

\begin{pf} 
	By (\ref{eq:cond0}), we have the following lower bound of the suboptimal cost $\hat{J}(t)$:
	\begin{align} \label{eq:lb_J}
	\hat{J}(t) &\geq	
	\underline{\gamma}_p(|\hat{x}(t-N|t) - z(t-N)|) \\
	& \ \ \  + \underline{\gamma}_w(|\hat{w}(i|t)|) + \underline{\gamma}_v(|\hat{v}(i|t)|) \notag
	\end{align}
	for all $i \in \mathbb{I}_{[t-N,t-1]}$.
	Note that the right-hand side in (\ref{eq:lb_J}) is greater than or equal to each of its three terms individually. By applying Lemma \ref{lem:boundedness_of_J}, it follows that
	\begin{align} 
	&|\hat{x}(t-N|t) - z(t-N)| 
	\leq \underline{\gamma}_p^{-1}(\hat{J}(t)) \notag\\
	& \stackrel{(\ref{eq:Kprop}),(\ref{eq:exp_bounds})}{\leq}   
	\underline{c}_p^{-1/a}  \Big( \bar{C}_p(N) |x_0 - \bar{x}_0|\rho^{t} \label{eq:bound_hat_x} \\
	& \ \ \
	+\bar{C}_w(N) \sum_{\tau=1}^{t} \rho^{\tau}|w(t-\tau)|
	+\bar{C}_v(N) \sum_{\tau=1}^{t} \rho^{\tau}|v(t-\tau)| \Big) \notag, 
	\end{align} 
	with constants $\bar{C}_p(N) := C_p (3\bar{c}\bar{\rho}_1(N))^{1/a}$, $\bar{C}_w(N) := C_w\rho^{-1}(3\bar{c}\bar{\rho}_2(N))^{1/a}$ and $\bar{C}_v(N) := C_v\rho^{-1}(3\bar{c}\bar{\rho}_2(N))^{1/a}$.	
	From (\ref{eq:lb_J}), a similar bound can be derived for $|w(i|t)|$ and $|v(i|t)|$ with $i \in \mathbb{I}_{[t-N,t-1]}$, and thus we obtain 
	\begin{align} 
		&|\hat{w}(i|t)| \leq 
		\underline{c}_w^{-1/a} \Big(\bar{C}_p(N) |x_0 - \bar{x}_0|\rho^{t} \label{eq:bound_hat_w} \\
		& \ \ \ \ \
		+ \bar{C}_w(N) \sum_{\tau=1}^{t} \rho^{\tau}|w(t-\tau)|	
		+ \bar{C}_v(N) \sum_{\tau=1}^{t} \rho^{\tau}|v(t-\tau)| \Big), \notag \\
		&|\hat{v}(i|t)| \leq
		\underline{c}_v^{-1/a} \Big(\bar{C}_p(N) |x_0 - \bar{x}_0|\rho^{t} \label{eq:bound_hat_v} \\
		& \ \ \ \ \
		+ \bar{C}_w(N) \sum_{\tau=1}^{t} \rho^{\tau}|w(t-\tau)|	
		+ \bar{C}_v(N) \sum_{\tau=1}^{t} \rho^{\tau}|v(t-\tau)|\Big). \notag
	\end{align}	
	Now, for all $t \in \mathbb{I}_{\geq N}$, consider the e-IOSS condition (\ref{eq:eIOSS}) with $x_1 = x(t-N)$, $x_2 = \hat{x}(t-N|t)$, $w_1(i) = w(i)$, $w_2(i) = \hat{w}(i|t)$, $h(x_1(i)) = y(i) - v(i)$ and $h(x_2(i)) = y(i) - \hat{v}(i|t)$ for $i \in \mathbb{I}_{[t-N,t-1]}$. Together with the triangle inequality it follows that
	\begin{subequations}\label{eq:proof_eta_all}
	\begin{align}
	| &x(t) - \hat{x}(t) | 
	\leq  c_p|x(t-N) - z(t-N)|\eta^N  \label{eq:proof_eta1} \\
	&  \ \ + c_p|\hat{x}(t-N|t) - z(t-N)|\eta^N \label{eq:proof_eta2}\\
	&  \ \ + c_w\sum_{\tau=1}^N\eta^{\tau}|w(t-\tau)|+ c_v\sum_{\tau=1}^N\eta^{\tau}|v(t-\tau)| \label{eq:proof_eta3} \\
	&  \ \ + c_w\sum_{\tau=1}^N\eta^{\tau}|\hat{w}(t-\tau|t)|  
	+  c_v\sum_{\tau=1}^N\eta^{\tau}|\hat{v}(t-\tau|t)| \label{eq:proof_eta4}
	\end{align}
	\end{subequations}
	for all $t \in \mathbb{I}_{\geq N}$. Now, the objective is to transfer (\ref{eq:proof_eta_all}) into the structure of (\ref{eq:RGES_result}).
	First, choose $\lambda := \max \lbrace \eta, \rho \rbrace$.  
	In (\ref{eq:proof_eta1}), we can apply the RGES property of the observer from Definition \ref{def:RGES}, i.e. (\ref{eq:RGES}) evaluated at time $t-N$. With $\lambda \geq \rho$ and $z_0=\bar{x}_0$, this yields
	\begin{align}
	&|x(t-N) - z(t-N)| \leq  \lambda^{-N}\Big(C_p|x_0 - \bar{x}_0|\lambda^{t} \label{eq:proof:eta_1} \\
	& \ \ \ \ \ \ 
	+ C_w\sum_{\tau=1}^t \lambda^{\tau}|w(t-\tau)| + C_v\sum_{\tau=1}^t \lambda^{\tau}|v(t-\tau)|\Big).  \notag
	\end{align}
	In (\ref{eq:proof_eta3}), since $\lambda \geq \eta$ and $t \geq N$, we have that
	\begin{equation}
	\sum_{\tau=1}^N\eta^{\tau}|w(t-\tau)| \leq \sum_{\tau=1}^{t} \lambda^{\tau}|w(t-\tau)|, \label{eq:proof:eta_2}
	\end{equation}
	which can also be applied to $\sum_{\tau=1}^N\eta^{\tau}|v(t-\tau)|$.
	Considering (\ref{eq:proof_eta4}), note that every single element $|\hat{w}(t-\tau|t)|$ and ${|\hat{v}(t-\tau|t)|}$ for all $\tau \in \mathbb{I}_{[1,N]}$ can be upper bounded by (\ref{eq:bound_hat_w}) and (\ref{eq:bound_hat_v}), respectively.
	Since these terms are independent of the variable $\tau$, we can pull them out of the sums in (\ref{eq:proof_eta4}).
	Then, by exploiting the geometric series, we can reduce the remaining sums according to
	\begin{equation} \label{eq:proof_ali}
	\sum_{\tau=1}^N\eta^{\tau} = \frac{\eta-\eta^{N+1}}{1-\eta} < \frac{\eta}{1-\eta} =: \bar{\eta}.
	\end{equation}	
	Now we apply (\ref{eq:proof:eta_1}) to (\ref{eq:proof_eta1}), (\ref{eq:bound_hat_x}) to (\ref{eq:proof_eta2}), (\ref{eq:proof:eta_2}) to (\ref{eq:proof_eta3}) and both (\ref{eq:bound_hat_w}) and (\ref{eq:bound_hat_v}) together with (\ref{eq:proof_ali}) to (\ref{eq:proof_eta4}), and we finally obtain (\ref{eq:RGES_result}) with $C_1,C_2,C_3$ replaced by
	\begin{align*}
	C_1'(N) := 
	&(c_p\eta^N\underline{c}_p^{-1/a} + c_w\bar{\eta}\underline{c}_w^{-1/a} 
	+ c_v\bar{\eta}\underline{c}_v^{-1/a})\bar{C}_p(N)\\
	&+ c_p(\eta/\lambda)^{N}C_p,\\
	C_2'(N) := 
	&(c_p\eta^N\underline{c}_p^{-1/a}
	+ c_w\bar{\eta}\underline{c}_w^{-1/a}
	+ c_v\bar{\eta}\underline{c}_v^{-1/a})\bar{C}_w(N)\\
	&+ c_p(\eta/\lambda)^{N}C_w 
	+ c_w,\\
	C_3'(N) := 
	& (c_p\eta^N\underline{c}_p^{-1/a}
	+ c_w\bar{\eta}\underline{c}_w^{-1/a} 
	+ c_v\bar{\eta}\underline{c}_v^{-1/a})\bar{C}_v(N)\\
	&+c_p(\eta/\lambda)^{N}C_v 
	+ c_v
	\end{align*}
	for all $t \in \mathbb{I}_{\geq N}$.
	It remains to show that this is also satisfied for $t \in \mathbb{I}_{[0,N-1]}$. 
	Since Lemma \ref{lem:boundedness_of_J} holds for all $t \in \mathbb{I}_{\geq0}$, it is straightforward to show that by applying Remark \ref{rem:emtpyN}, (\ref{eq:proof_eta1})-(\ref{eq:proof_eta4}) also hold for $t \in \mathbb{I}_{[0,N-1]}$ with $N$ replaced by $t$.
	Because $x(0)-z_0 = x_0 - \bar{x}_0$, in (\ref{eq:proof_eta1}) we do not need to consider the observer this time. 
	Since $\eta^t\leq 1$ and $\sum_{\tau=1}^{t}\eta^{\tau} <\bar{\eta}$ for all $t \in \mathbb{I}_{[0,N-1]}$, we obtain (\ref{eq:RGES_result}) with $C_1,C_2,C_3$ replaced by
	\begin{align*} 
	C_1''(N) &:= 
	c_p
	+ (c_p\underline{c}_p^{-1/a} 
	+ c_w\bar{\eta}\underline{c}_w^{-1/a}
	+ c_v\bar{\eta}\underline{c}_v^{-1/a})\bar{C}_p(N),\\ 
	C_2''(N) &:=  c_w 
	+ (c_p\underline{c}_p^{-1/a} 
	+ c_w\bar{\eta}\underline{c}_w^{-1/a} 
	+ c_v\bar{\eta}\underline{c}_v^{-1/a})\bar{C}_w(N),\\
	C_3''(N) &:= c_v 
	+ (c_p\underline{c}_p^{-1/a} 
	+ c_w\bar{\eta}\underline{c}_w^{-1/a}
	+ c_v\bar{\eta}\underline{c}_v^{-1/a})\bar{C}_v(N).
	\end{align*}
	Now, with $C_i := \max \lbrace C_i',C_i'' \rbrace$ for every $i = \lbrace 1,2,3 \rbrace$, observe that (\ref{eq:RGES_result}) holds for all $t \in \mathbb{I}_{\geq0}$ and thus the suboptimal estimator is RGES as desired. \qed
\end{pf}

\begin{remark}
	Note that RGES of the proposed suboptimal estimator is guaranteed for all $N \in \mathbb{I}_{\geq1}$. In other words, there is no minimum required horizon length $N_0$, as it was the case, e.g., in \cite{Mueller2017,Hu2017,Knuefer2018,Allan2019a}. 
	This is due to the fact that we do not require contraction of the estimation error from time $t-N$ to $t$, but instead establish  stability directly by the use of an additional RGES observer.
\end{remark}	



\section{SIMULATION EXAMPLE} \label{sec:simulation}
In order to illustrate our results, we consider the reaction $2A\rightleftharpoons B$ taking place in a constant-volume batch reactor \citep[][]{Tenny2002}, which can be described as
\begin{equation} \label{ex:eq:sys}
\dot{x} = 
\begin{pmatrix} -2k_1x_1^2 + 2k_2x_2 \\ k_1x_1^2 - k_2x_2 \end{pmatrix}, \quad	y = x_1+x_2 
\end{equation} 
with $k_1 = 0.16$, $k_2 = 0.64$, and $x_0 = \begin{pmatrix} 5 & 2 \end{pmatrix}^T$.
For system~(\ref{ex:eq:sys}), a Luenberger-like nonlinear observer can be designed as 
\begin{equation} \label{ex:eq:obs}
\dot{z} = 
\begin{pmatrix} -2k_1z_1^2 + 2k_2z_2 \\ k_1z_1^2 - k_2z_2 \end{pmatrix}
+ \begin{pmatrix} L_1 \\ L_2 \end{pmatrix} (y-z_1-z_2)
\end{equation} 
with $L_1 = L_2 = 0.5$ and $z_0 = \begin{pmatrix} 3 & 0 \end{pmatrix}^T$.
In the simulation, we discretize system (\ref{ex:eq:sys}) and observer (\ref{ex:eq:obs}) with the sampling time $\Delta = 0.1$.
We assume that the system is subject to random disturbances $w \sim \mathcal{N}(0,Q)$ and $v \sim \mathcal{N}(0,R)$ with $Q = 0.1^2I_2$ and $R=0.2^2$, respectively.
We design the proposed suboptimal estimator using a horizon length of $N=10$, quadratic costs $l(\omega,\nu) = \omega^TQ^{-1}\omega + \nu^TR^{-1}\nu$ and $\Gamma(\chi,z(t-N)) = (\chi- z(t-N))^T (\chi- z(t-N))$.
The simulation was performed using the software tools CasADi \citep{Andersson2018} and the integrated solver Ipopt \citep{Waechter2005}; the optimization algorithm was terminated after $i = 0,2,5$ iterations.
\begin{figure}[t]
	\centering
	\includegraphics{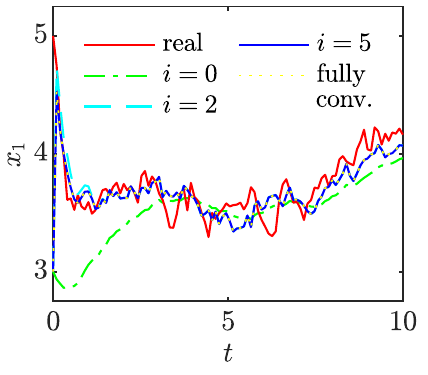}
	\includegraphics{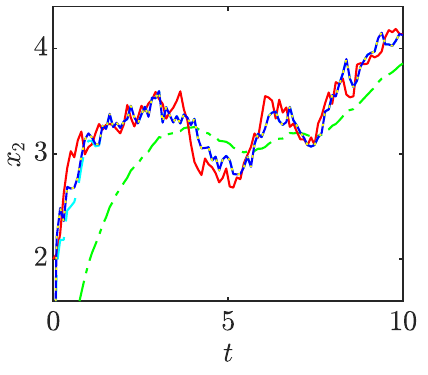}
	\caption{Comparison between suboptimal MHE with $i = 0,2,5$ iterations, fully converged MHE, and real system state.\vspace{3ex}}
	\label{ex:figure}
\end{figure}
Fig.~\ref{ex:figure} shows the simulation results and reveals that the proposed suboptimal estimator is RGES independently of the number of iterations $i$.
For $i=0$, the suboptimal estimator follows the trajectory of  observer (\ref{ex:eq:obs}) since no optimization is performed. Because the observer was designed rather conservatively, the corresponding estimates converge slowly to the true state. 
Although this trajectory is used as a (rather poor) warm start for the optimization problem, already only $i=2$ iterations yield a much better state estimate, and after $i=5$ iterations, the estimated trajectory becomes indistinguishable compared to fully converged MHE.
In summary, we find that suboptimal MHE can improve the state estimate of a (potentially poorly designed) auxiliary observer already with (very) few iterations, while guaranteeing robust stability independent of the number of iterations as guaranteed by Theorem~\ref{theorem}.

\section{CONCLUSIONS} \label{sec:conclusions}
In this paper, we presented a suboptimal moving horizon estimator and established robust stability by requiring that a suboptimal solution improves the cost of a given candidate solution, similar as in MPC.
In particular, robust stability of the proposed suboptimal moving horizon estimator is guaranteed independent of the horizon length and the number of solver iterations. Furthermore, our simulation results illustrate that already (very) few iterates can significantly improve the state estimate compared to the auxiliary observer used to construct the candidate solution.
In our current work, we aim to relax the assumption that the auxiliary observer needs to be globally stable by re-initializing the observer at each time step at the beginning of the horizon  \cite[cf.][]{Sui2010, Suwantong2014, Liu2013, Gharbi2020}. 
We also aim to consider more general classes of nonlinear systems and, in addition, provide guarantees that become better with an increasing estimation horizon.

{\scriptsize
	\bibliography{/home/schiller/Seafile/IRT/Literatur/Literatur.bib}
}                                                                 
\end{document}